\documentclass[final,5p,floatfix,floats,superscriptaddress,times,twocolumn,showpacs,amsmath,amssymb]{elsarticle}

\usepackage{graphicx}
\usepackage{bm}
\usepackage{ulem}
\usepackage{epsfig}
\usepackage{amssymb}
\usepackage{amsthm}

\def\vec#1{{\bf#1}}
\def\eq#1{Eq.\ (\ref{#1})}

\def\fig#1{Figure~\ \ref{#1}}

\journal{Physica E}

\begin{document}

\begin{frontmatter}



\title{Singlet-triplet transitions in highly correlated nanowire quantum dots}


\author[1]{Y. T. Chen}
\address[1]{Department of Electrophysics, National Chiao Tung
  University, Hsinchu 30010, Taiwan, Republic of China}
\author[1]{C. C. Chao}
\author[1]{S. Y. Huang}
\author[2]{C. S. Tang}
\address[2]{Department of Mechanical Engineering,
        National United University, 1, Lienda, Miaoli 36003, Taiwan, Republic of China}
\author[1]{S. J. Cheng\corref{cor1}}
\ead{sjcheng@mail.nctu.edu.tw}

\cortext[cor1]{Corresponding author}

\begin{abstract}
We consider a quantum dot embedded in a three-dimensional nanowire with
tunable aspect ratio $a$. A configuration interaction theory is
developed to calculate the energy spectra of the finite 1D quantum dot
systems charged with two electrons in the presence of magnetic fields
$B$ along the wire axis.  Fruitful singlet-triplet transition behaviors
are revealed and explained in terms of the competing exchange interaction,
correlation interaction, and spin Zeeman energy. In the
high aspect ratio regime, the singlet-triplet transitions are shown
designable by tuning the parameters $a$ and $B$. The transitions also manifest the
highly correlated nature of long nanowire quantum dots.
\end{abstract}


\end{frontmatter}


\label{}

\section{Introduction}

For years, few electron charged quantum dots have attracted extensive
attention due to the controllable electronic and spin properties
~\cite{Reimann02}. However, only few attempts have so far been made for
studies of finite 1D nanowire quantum dots (NWQDs). More recently, it
was shown that the NWQDs formed in the heterostructures in nanowires
can be fabricated as single electron transistors and successively
charged with controlled number of electrons~\cite{Bjork04}.  The
successful experimental works motivate us to explore the possible
geometric effects of NWQDs characterized by their aspect ratios, $a$,
on the electronic and spin properties of two-electron charged NWQDs.

In this work, we focus on the study of the singlet-triplet (ST)
transitions in two-electron charged NWQDs~\cite{Fasth07}, conducted by using a
developed configuration interaction (CI) theory in combination with the
exact diagonalization techniques based on a 3D asymmetric parabolic
model. It will be illustrated that the ST transitions in
InAs-based NWQDs driven by an appropriate magnetic field are associated
with the competing effects of large spin-Zeeman energies as well as the
exchange and correlation energies. The correlation-dominated nature
of a long NWQD (i.e. with high aspect ratio) will be identified by the spin phase
diagram with respect to the applied magnetic fields and the tunable
aspect ratio.

\section{Theoretical Model}
\subsection{single-electron spectrum}

We begin with the single electron problem of a NWQD with axial magnetic field
${\bf B}=(0,0,B)$, described by the Hamiltonian~\cite{Chen06}
\begin{equation}
H_0 =\frac{1}{2m^\ast}(\vec{p}+e\vec{A})^2 + V(x,y,z) + H_{\rm Z}\, ,
\label{H_sp}
\end{equation}
where ${\bf A}=(B/2)(y,-x,0)$ denotes the vector potential and $m^\ast$
stands for the effective mass of an electron with charge $-e$.  The
spin-Zeeman Hamiltonian $H_{\rm Z} =  g^\ast \mu_{B}B s_z$ is in terms
of the z-component of electron spin $s_z$ and the effective Lande
g-factor of electron $g^\ast$ and the Bohr magneton $\mu_B$.  In
addition, the confining potential $V(x,y,z)=m^\ast \left[\omega_0^2 \left(x^2 + y^2 \right)
+\omega_z^2 z^2 \right]/2$ is assumed of the parabolic form with $\omega_0$ and $\omega_z$
parametrizing, respectively, the transverse and the longitudinal
confining strength.

In this work, we assume a constant $g^\ast$  and take the $g^\ast=-8.0$
for InAs~\cite{Fasth07,Bjork05}. The single electron Hamiltonian (\ref{H_sp}) leads to the
extended Fock-Darwin single-particle spectrum
\begin{eqnarray}
\epsilon_{n,m,q,s_z}&=& \hbar\omega_{+}\left(n+\frac{1}{2}\right) +
\hbar\omega_{-}\left(m+\frac{1}{2}\right) \nonumber \\
&& + \hbar\omega_{z}\left(q+\frac{1}{2}\right) + E_{\rm Z}
 \label{spspec}
\end{eqnarray}
where $n,m,q=0,1,2\cdots$ denote oscillator quantum numbers, $s_z=+\frac{1}{2}$ ($s_z=-\frac{1}{2}$) the projection of electron spin $\uparrow$ ($\downarrow$)
$E_{\rm Z}= g^\ast \mu_{B}B s_z$ the spin Zeeman energy, and $\omega_{\pm}=\omega_h \pm \omega_c/2$
is defined in terms of the hybridized frequency $\omega_h\equiv
(\omega_0^2+\omega_c^2/4)^{1/2}$ and  the cyclotron frequency $\omega_c={eB}/{m^\ast}$.
.

The eigenstate $|n,m,q\rangle$ possesses the orbital angular momentum
$l_z=\hbar (n-m)$ and the parity $P=1$ ($P=-1$) with respect to
$z-$axis for the even (odd) $q$ number.  The wave function of the
lowest orbital is given by $\psi_{000}(\vec{r})= \exp \left[ -\left(\left(x^2+y^2\right)/{l_h^2} + {z^2}/{l_z^2} \right)/4 \right]
/\left(2 \pi^{3/4} {l_h} \sqrt{l_z}\right)$
with the characteristic lengths of the wave function extents
$l_h=\sqrt{{\hbar}/{2 m^\ast \omega_h}}$ and $l_z=\sqrt{{\hbar}/{2
m^\ast \omega_z}}$, from which one can generate the wave functions of
any other excited states by successively applying raising operators
\cite{Hawrylak93}. For most synthesized NWQDs, the diameter of the
cross section is of the scales $50$~nm, while the length of wire could
be tunable over a wide range from $10$~nm to $300$~nm~ \cite{Bjork05}.
To characterize the geometry of NWQDs, we define the aspect ratio
parameter,
\begin{equation}
a\equiv
\frac{l_z}{l_0}=\sqrt{\frac{\omega_0}{\omega_z}}\sim\frac{L_z}{L_x}\, .
\end{equation}
according to the extents of the wave function.
\begin{figure}
\includegraphics[height=5cm]{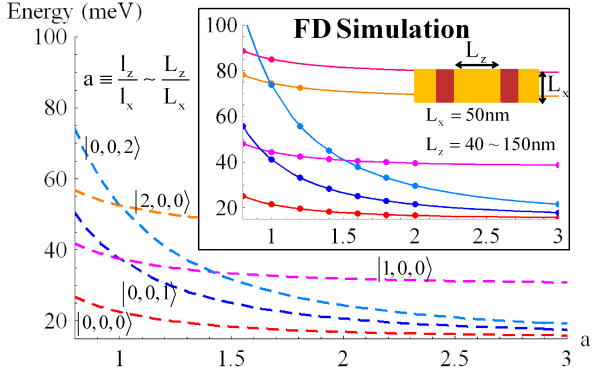}
\caption{Single-electron energy spectrum as a function of aspect ratio
$a$ with no magnetic field. The dot diameter $L_x$=$50$~nm and the dot
height $L_z$ varies from $40\rm{nm}$ to $150\rm{nm}$.  The finite
difference results are shown in the inset for comparison.}
\label{para_FD}
\end{figure}

\begin{figure}
\includegraphics[width=7 cm]{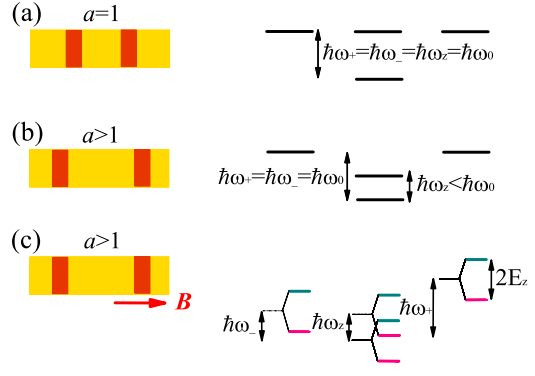}
\caption{Schematic diagram of the lowest two orbitals occupied by two
relevant electrons for the cases: (a) isotropic ($a=1$) and $B=0$; (b)
``rod-like'' ($a>1$) and $B=0$; (c) with nonzero electronic $g$-factor
and $B\ne 0$. $2E_Z= g^\ast\mu_BB$ indicates the spin Zeeman energy
splitting.} \label{sys_illus}
\end{figure}

Figure \ref{para_FD} shows the single electron energy spectra, as a
function of $a$, of NWQDs at zero magnetic field.
 To examine the
validity of the model,  we carry out a numerical finite difference (FD)
simulation for the electronic structure of InAs/InP heterostructure
NWQD, as shown in the inset of \fig{para_FD}. The InAs NWQD is embedded
in InP barriers with the diameter $L_x=50$~nm and varying the length
from $L_z=40$~nm to $L_z=150$~nm. The effective mass $m^\ast=0.023m_0$
and the barrier offset $V_b=0.6$~eV are taken~\cite{Bjork02}. The
confining strength parameter $\hbar \omega_0$ is fit by the ground
state energy from the FD simulation. We set $\hbar \omega_0=13.3$~meV
for $L_x=50$~nm NWQDs throughout this paper. The numerically calculated
energy spectrum is in good agreement with that given by parabolic
model. The schematic illustration of the engineered single electron
energy levels of NWQDs by tunable $a$ and applied $B$ is shown in
\fig{sys_illus}.

\subsection{Interacting NWQD}

To investigate the few-electron interaction effects in a NWQD, we express the few electron Hamiltonian in second quantization as
\begin{eqnarray}
H &=& \sum_{i,\sigma}\epsilon_{i\sigma}
c_{i\sigma}^{\dag}c_{i\sigma}\nonumber \\
&&+ \frac{1}{2}\sum_{ijkl,\sigma \sigma'}\langle ij|V|kl\rangle
c_{i\sigma}^{\dag}c_{j\sigma'}^{\dag}c_{k\sigma'}c_{l\sigma} \, ,
\label{full_H}
\end{eqnarray}
where $i,j,k,l$ stand for the composite indices of single electron
orbitals (e.g. $|i\rangle=|n_i,m_i,q_i\rangle$), $\sigma=\uparrow/\downarrow$ denotes the electron spin with $s_z=+\frac{1}{2}/-\frac{1}{2}$, and $c_{i\sigma}^{\dag}$ ($c_{i\sigma}$)
is the electron creation (annihilation) operators.  The first (second)
term on the right hand side of Eq.(\ref{full_H}) represents the kinetic
energy of electrons (the Coulomb interactions between electrons) and
the Coulomb matrix elements are defined as $\langle ij|V|kl\rangle \equiv e^2\left(4\pi\kappa\right)^{-1}  \int \int
d\vec{r_1} d\vec{r_2}\psi_i^*(\vec{r_1})\psi_j^*(\vec{r_2})
 \left(|\vec{r_1} - \vec{r_2}|\right)^{-1}
 \psi_k(\vec{r_2})\psi_l(\vec{r_1})$, where $\kappa$ is the dielectric constant of dot material (
$\kappa=15.15\epsilon_0$ is taken for InAs throughout this work). After
lengthy derivation, for NWQDs with $a\ge 1$, we obtain the following
formulation of the Coulomb matrix elements:
\begin{eqnarray}
&&\langle
n_{i}m_{i}q_{i};n_{j}m_{j}q_{j}|V|n_{k}m_{k}q_{k};n_{l}m_{l}q_{l}\rangle
\label{Vijkl1}
\\
&=& (\frac{1}{\pi l_h})   \frac{ \delta_{R_L,R_R} \cdot
\delta_{q_i+q_j+q_l+q_k, \rm{even}}
  }{\sqrt{n_i!m_i!q_i!n_j!m_j!q_j!n_k!m_k!q_k!n_l!m_l!q_l!}} \nonumber\\
&\times&  \sum_{p_1=0}^{\min(n_i,n_l)} \sum_{p_2=0}^{\min(m_i,m_l)}
\sum_{p_3=0}^{\min(q_i,q_l)} \sum_{p_4=0}^{\min(n_j,n_k)}
\sum_{p_5=0}^{\min(m_j,m_k)} \sum_{p_6=0}^{\min(q_j,q_k)}\nonumber \\
&\times& p_1!p_2!p_3!p_4!p_5!p_6! \nonumber \\
&\times& {n_i \choose p_1} {n_l \choose p_1} {m_i \choose p_2} {m_l
\choose p_2} {q_i \choose p_3} {q_l \choose p_3} \nonumber \\
 &\times& {n_j \choose p_4} {n_k \choose p_4} {m_j \choose p_5} {m_k \choose p_5}
{q_j \choose p_6}
{q_k \choose p_6}\nonumber\\
&\times&  (-1)^{u+v/2+n_j+m_j+q_j+n_k+m_k+q_k} \times (\frac{1}{2})^u \times x^{u+1/2}\nonumber\\
&\times&
\frac{\Gamma(\frac{1+2u+v}{2})\Gamma(1+u)\Gamma(\frac{1+v}{2})}{\Gamma(\frac{3+2u+v}{2})} \nonumber \\
&\times&  _2F_1(1+u,\frac{1+2u+v}{2};\frac{3+2u+v}{2};1-x)\, ,\nonumber
\label{Vijkl}
\end{eqnarray}
where we have defined $u=m_i+m_j+n_l+n_k-(p_1+p_2+p_4+p_5)$,
  $v=(q_i+q_l+q_j+q_k)-2(p_3+p_6)$,
  $R_L=(m_i+m_j)-(n_i+n_j)=-(L_{z,i}+L_{z,j})$,
  $R_R=(m_l+m_k)-(n_l+n_k)=-(L_{z,l}+L_{z,k})$,
$x\equiv \omega_z/\omega_h$, and $_2F_1$ is the hypergeometric function. The $\delta$-functions
$\delta_{q_i+q_j+q_l+q_k, \rm{even}}$ and $\delta_{R_L,R_R}$ in the
formulation ensure the conservation of the parity with respect to
$z$-axis and the $z$-component of angular momentum of system $L_z$,
respectively. The formulation of \eq{Vijkl1} is reexamined by computing
the Coulomb integral numerically.

\subsection{Exact diagonalization}

The energy spectrum of an interacting two-electron NWQD is
calculated using the standard numerical exact diagonalization technique
\cite{Hawrylak03}. The numerically {\it exact} results are obtained by
increasing the numbers of chosen single electron orbital basis and the
corresponding two-electron configurations until a numerical
convergence is achieved. In our full configuration interaction (FCI)
calculations, we chose the typical orbital number from $20$ to $26$ and
the number of corresponding configurations is from $190$ to $325$. In
order to highlight the Coulomb correlations, we also carry out the
partial CI (PCI) calculations in which only the lowest energy $N_e$
configuration is taken and compare the PCI results with those obtained
from the FCI calculations.

\section{Results and discussion}

\begin{figure}
\includegraphics[height=5.5cm]{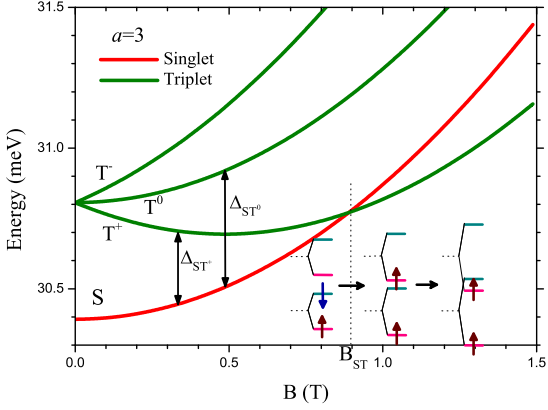}
\caption{Correlated two-electron energy spectrum as a function of
magnetic field in a NWQD with diameter $L_x = 50$~nm and aspect ratio
$a=3$.} \label{2eEB}
\end{figure}
Figure \ref{2eEB} presents the FCI result of magneto-energy spectrum of
two interacting electrons in a NWQD with $a=3$. The ST transition of
the two-electron ground state is shown to happen as $B_{\rm ST}\sim
0.9$~T. As the applied magnetic field is weak, the spin Zeeman splitting is small and the two electrons mostly doubly fill the lowest S-orbital.
With increasing magnetic field increases, the energy difference between triplet and singlet states of the two electrons decreases because of increasing spin Zeeman and exchange energies, both of which energetically favor the triplet states $|T^+\rangle$.
As the applied  field is higher than $B_c\sim 0.9$T, the ground state of two electrons transit from the singlet state to the triplet one.

\begin{figure}
\includegraphics[height=4.22cm]{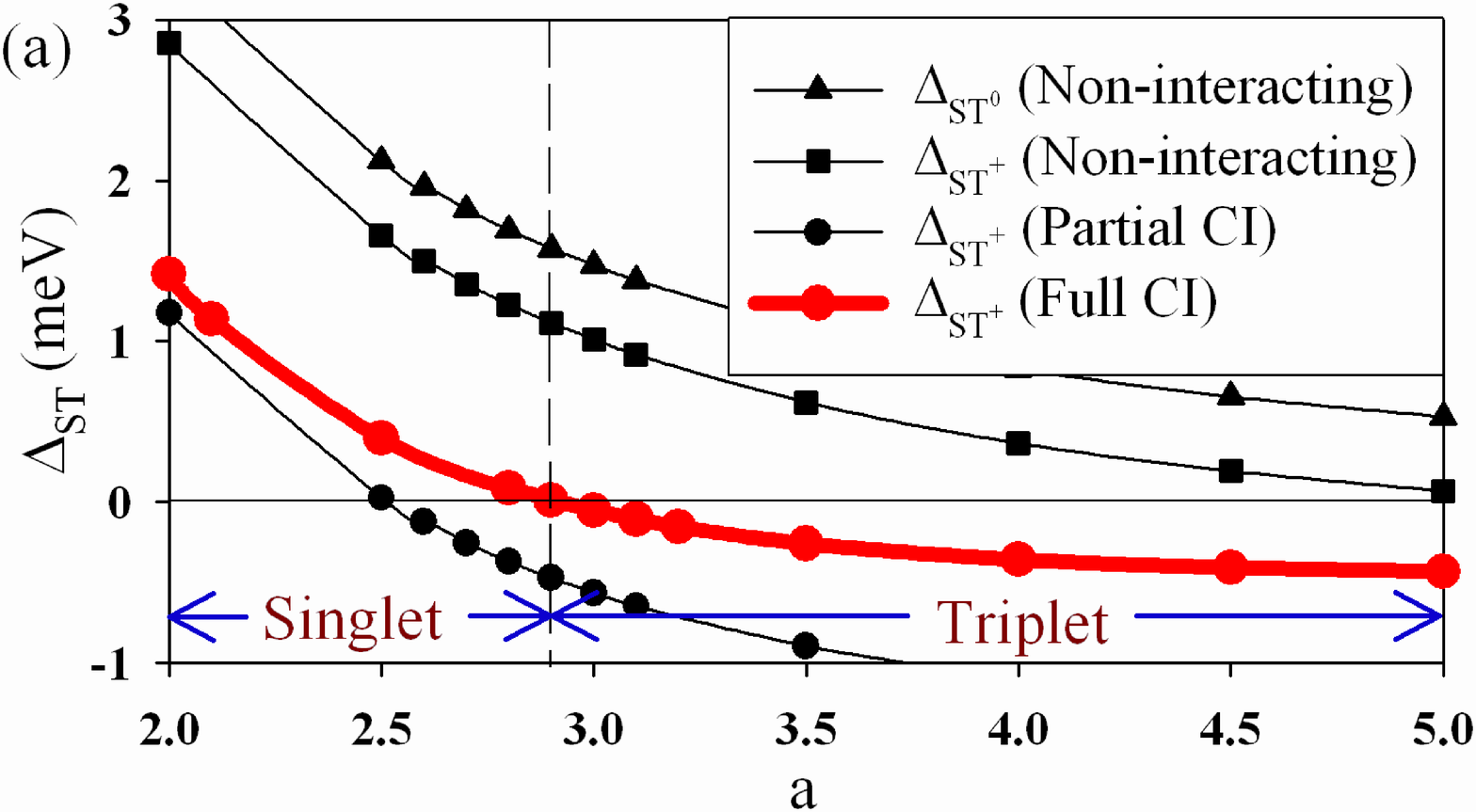}
\includegraphics[height=3.5cm]{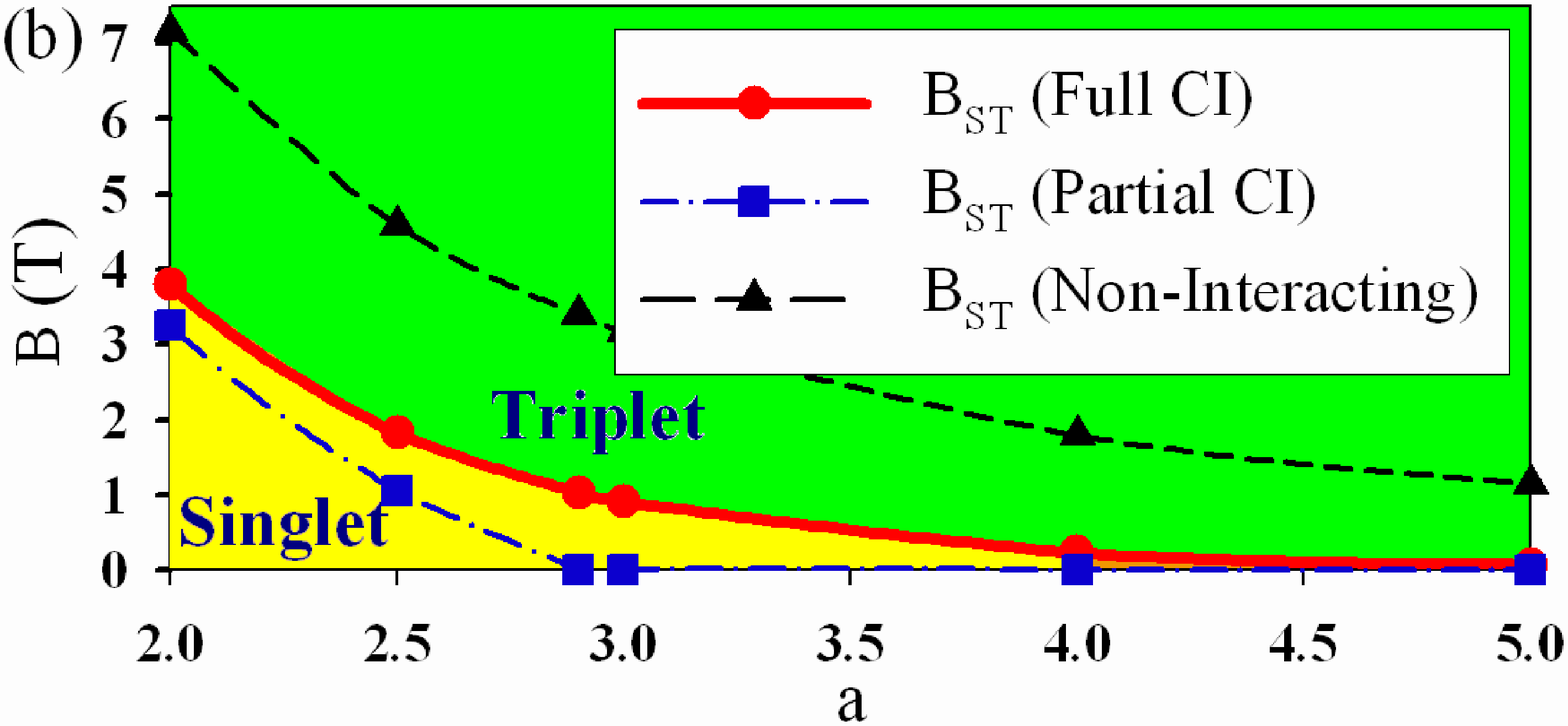}
\caption{(a) Singlet-triplet splitting $\Delta_{{\rm ST}^i}$ as a
function of aspect ratio $a$ with $B=1$~T. (b) The spin phase diagram
for electrons making singlet-triplet transition with respect to
magnetic field $B$ versus aspect ratio $a$.} \label{DST-BST}
\end{figure}
Now we turn to study the spin singlet-triplet splitting as a function of $a$, defined by $\Delta_{{\rm ST}^i}\equiv E_{T^i} - E_S$,
with $i=-$, $0$, and $+$ corresponding to the $T^-$, $T^0$, and $T^+$
triplet states, respectively. In Figure \ref{DST-BST}(a), we show the
$\Delta_{{\rm ST}^i}$ as a function of aspect ratio $a$ under a fixed
magnetic field $B=1$~T. In the non-interacting case, $\Delta^0_{\rm
ST^i}$ are shown to decrease monotonically with increasing  aspect
ratio $a$. Since only $T^+$ energy is decreased by spin-Zeeman term,
the ST transition could only occur between $S$ and $T^+$ states. That
is, only $\Delta_{{\rm ST}^+}$ crosses zero as $a$ is very large, while
$\Delta_{{\rm ST}^0}$ remains positive always. Thus below we shall only
consider $\Delta_{{\rm ST}^+}$ for the discussion of ST transition. The
non-interacting ST splitting can be derived as $\Delta^0_{\rm ST^+} =
\hbar\omega_0/a^2 - 2E_{\rm Z}$, explicitly showing the quadratic
decrease of $\Delta^0_{\rm ST^+}$ with respect to $a$. Accordingly, in
the non-interacting picture, the  critical aspect ratio $a_{\rm ST} $
where the ST transition occurs is predicted as $a_{\rm ST} =
\sqrt{\hbar \omega_0 /g^* \mu_B B}$.

However, the PCI calculation predicts a much smaller value of critical aspect ratio $a_{\rm
ST} = 2.5$.
 In the PCI result, the ST splitting is substantially reduced by the energy reduction of the T state due to the reduced direct Coulomb interaction and the negative exchange interaction between the two electrons in the state.
The FCI calculation shows $a_{\rm ST} = 2.9$, as indicated by the dashed
vertical line in Figure \ref{DST-BST}(a). In fact, the difference in the values of $\Delta_{\rm ST^+}$ obtained from the PCI and FCI calculations increases as $a$ increases. This indicates that the Coulomb correlation effect tends to increase the ST splitting again and becomes even more pronounced in long NWQD with high $a$.

Figure \ref{DST-BST}(b) shows the calculated spin phase diagram of two-electron NWQDs with respect to the aspect ratio $a$ and applied magnetic field $B$. The spin singlet and triplet phases, appearing in the low $a$-$B$ and high $a$-$B$ regimes, respectively, are distinguished by the curve of $B_{\rm ST}$ which show a monotonic decrease with $a$.
For noninteracting electrons, the
critical magnetic field can be derived as $B_{\rm ST}=\hbar\omega_0/g\mu_B a^2$, showing a quadratic decay with $a$.

In comparison with the non-interacting cases, the PCI calculations obtain the $B_{\rm ST}$ that is significantly reduced and goes to zero for $a>2.9$. In the one-configuration approximation used in the PCI calculation, the ST
splitting is given by $\Delta_{\rm ST^+}\approx \Delta^0_{\rm ST^+} + \Delta_{\rm
ST}^{\rm dir} - V_{\rm T}^{\rm ex}$, where $\Delta^0_{\rm ST^+}$ is the ST splitting in the non-interacting cases, $\Delta_{\rm ST}^{\rm dir}
\equiv V_{\rm T}^{\rm dir}-V_{\rm S}^{\rm dir}< 0$ is the direct
energy difference between the triplet and the singlet states, and
$V_{\rm T}^{\rm ex}$ is the exchange energy between electrons in the
$T^+$ state. Accordingly, we obtain $B_{\rm ST} = (\hbar\omega_0/a^2 + \Delta_{\rm ST}^{\rm dir} - V_{\rm T}^{\rm ex}) / g^\ast \mu_B$.  In the large
aspect ratio regime, the negative $\Delta_{\rm ST}^{\rm dir}$ and $V_{\rm T}^{\rm ex}$ reduce
the $E_{{\rm T}^+}$ and $B_{\rm ST}=0$ results for $a>3$.
 However, the FCI calculation predict larger and always positive $B_{ST}$. In fact, as increasing $a$, the relative strength of electronic Coulomb correlations increases because of reduced $\hbar \omega_z$ and strong configuration interactions. Such $a$-engineered Coulomb correlations energetically favor the singlet two-electron states and result in the non-zero $B_{ST}$ in the high aspect ratio regime.

\section{Summary}

In conclusion, a configuration interaction (CI)
theory is developed for studying the magneto-energy spectra and the singlet-triplet transitions of two-electron NWQDs with longitudinal magnetic field $B$ and tunable aspect ratio $a$. For short NWQDs of low aspect ratio $a<3$, the ST transition behaviors are dominated by the spin Zeeman, Coulomb direct and  exchange energies, and can be well studied by using PCI calculation. However, our FCI calculations show the increasing importance of Coulomb correlations in long NWQDs with increasing aspect ratio $a$ over $3$. The FCI calculation present the spin phase
diagram of a two-electron NWQD which are highly dependent on $a$, and suggests the controllability of singlet or triplet spin states by changing the aspect ratio of NWQD.

\section{Acknowledgment}

This work was supported in part by the National Science Council of the
Republic of China through Contracts No. NSC95-2112-M-009-033-MY3 and
No. NSC97-2112-M-239-003-MY3.



\end{document}